\documentclass{article}

\usepackage{arxiv}

\usepackage[utf8]{inputenc} 
\usepackage[T1]{fontenc}    
\usepackage{hyperref}       
\usepackage{url}            
\usepackage{booktabs}       
\usepackage{amsfonts}       
\usepackage{nicefrac}       
\usepackage{microtype}      
\usepackage{lipsum}
\usepackage{graphicx}
\usepackage{todonotes}
\usepackage{soul}
\usepackage{amsmath}
\usepackage{empheq}
\usepackage{float}
\usepackage{graphicx}
\usepackage{caption}
\usepackage{subcaption}
\usepackage[utf8]{inputenc}
\usepackage[T1]{fontenc}
\graphicspath{ {./images/} }

\title{Thermal Noise Magnetometry of magneticnanoparticles: simulations and experiments}

\author{
 Katrijn Everaert \\
  Physikalisch-Technische Bundesanstalt\\
  Berlin 10587, Germany\\
  Department of Solid State Sciences\\
  Ghent University\\
  9000 Ghent, Belgium\\
  \texttt{katrijn.everaert@ptb.de} \\
  \texttt{katrijn.everaert@ugent.be} \\
   \And
 Maik Liebl \\
  Physikalisch-Technische Bundesanstalt\\
  Berlin 10587, Germany\\
  \And
 Dirk Gutkelch \\
  Physikalisch-Technische Bundesanstalt\\
  Berlin 10587, Germany\\
    \And
 Bartel Van Waeyenberge \\
  Department of Solid State Sciences\\
  Ghent University\\
  9000 Ghent, Belgium\\
    \And
 Frank Wiekhorst \\
  Physikalisch-Technische Bundesanstalt\\
  Berlin 10587, Germany\\
    \And
 Jonathan Leliaert \\
  Department of Solid State Sciences\\
  Ghent University\\
  9000 Ghent, Belgium\\
}
\date{}
\begin{document}
\maketitle
\begin{abstract}
Magnetic nanoparticles have proven to be valuable for biomedical applications. A prerequisite for the efficiency of these applications are precisely characterized particles. Thermal Noise Magnetrometry (TNM), unlike any other magnetic characterization method, allows to characterize the particles without the application of an external excitation. We present a theoretical framework to model the experiment in order to examine the magnetic noise power properties of particle ensembles in TNM and optimize the geometry of the nanoparticle sample. An optimized sample geometry is calculated to maximize the noise power measured by the detector. The theoretical framework and the newly designed sample holder are validated by measuring a sensitivity profile in the setup. The optimized sample holder increases the noise power with a factor of 3.5 compared to the regular sample holder, while its volume is decreased by more than half. Moreover, the experimental profiles are consistent with the simulated counterparts, confirming the relevance of the theoretical framework. These results contribute to the further establishment of TNM as magnetic nanoparticle characterization method. 
\end{abstract}


\section{Introduction}
Thermal fluctuations are ubiquitous in a broad range of physical systems. Although they are often unwanted and referred to as noise, they also carry information. Einstein was the first to model the movements of colloidal particles in a fluid \cite{Einstein1905}, more commonly known as Brownian motion, which is for instance used in the application ``Dynamic Light Scattering'' (DLS). This is a powerful characterisation technique for macromolecular systems which working principles rest on the spontaneous movement of the macromolecules in their suspension \cite{Berne1976,Stetefeld2016}. 
Also the thermal agitation of electrons in a conductor (Johnson noise), measured and described by Johnson and Nyquist\cite{Johnson1928,Nyquist1928}, has applications, e.g., Johnson noise thermometry\cite{White1996}, which is especially relevant in low-temperature systems because it operates without a driving current and heat dissipation is minimal\cite{Beyer2007,Rothfuss2016}. A general theory about thermal noise led to the fluctuation-dissipation theorem (FDT) \cite{Callen1951,Khatami1966}: the fluctuations in any extensive quantity are related with the dissipation in the system as a result of the application of its conjugate intensive quantity\cite{Refregier2004}. Although some of these descriptions go back decades in time, the study of thermal fluctuations and the FDT today are an active research field with several applications in various disciplines. The FDT has for instance also been extended to non-equilibrium and non-Hamiltonian systems\cite{Marini2008}, allowing one to e.g. model the response of a climate system to a weak external forcing \cite{Leith1975,Gritsun2007,Cooper2011}.\\

In this paper, we will consider magnetic nanoparticle  (MNP) dynamics, which are inherently stochastic due to their nanoscopic sizes. The MNPs typically consist of a magnetic core (e.g. ferrite) and a non-magnetic shell (e.g. silica) and over the past decades have found their way into numerous biomedical applications\cite{Pankhurst2003,TON-19}. They can be used in imaging, either directly, as in magnetic particle imaging\cite{GLE-05,knopp2012magnetic} (MPI) or Magnetorelaxometry imaging\cite{Liebl2015,coene2017multi} (MRXi), or indirectly as contrast agent\cite{laurent2016mri} in magnetic resonance imaging (MRI). They can serve as heat generators in cancer treatment (magnetic nanoparticle hyperthermia)\cite{etemadi2020magnetic,ESP-16} and can be used in targeted drug delivery\cite{al2016magnetic}.  
For these applications to work safely and reliably, there is a need for accurate MNP characterization techniques. Among others\cite{WEL-17}, Magnetorelaxometry\cite{Wiekhorst2012} (MRX) and AC-Susceptibility\cite{Ludwig2010} (ACS) measurements are well-established and are capable of determining, e.g. both the core size and the hydrodynamic size of the particles.\\

In MRX, the relaxation of the net magnetic moment of a nanoparticle sample is recorded, after abruptly switching off the external magnetic field. In ACS, the magnetic susceptibility is measured, making it e.g. able to determine the particle size distribution from the phase lag between the applied sinusoidal magnetic field and the magnetic moments of the particles\cite{Dieckhoff2016}. It is worth noting that both MRX and ACS (similar to all direct magnetic characterization techniques\cite{Leliaert2015}) record the particles' response to an externally applied field, and are therefore methods based on the \textit{dissipation} part of the FDT: the dissipation or impedance of the magnetization $M$ of the sample, which is an extensive quantity, is studied as a result of the application of the externally applied field $H$, the intensive quantity. The thermal \textit{fluctuations} in $M$ in the absence of any external field are thus caused by the very same mechanisms and can thus also be used to characterize nanoparticle systems\cite{Leliaert2017}. The technique of Thermal Noise Magnetometry (TNM), introduced by Leliaert \textit{et al} in Ref. \cite{Leliaert2015}, could be a powerful alternative for the characterization of magnetic nanoparticles dispersions, with diminutive impact on the sample: unlike any other characterisation method, no external field excitations are applied which could potentially alter the particles magnetization state\cite{Bharti2015,myrovali2016arrangement}.\\

On the timescale of typical MNP characterisation measurements, the dynamics of the MNPs are driven by two different stochastic processes.
Together, they give rise to both the relaxations observed in MRX and the fluctuations considered in TNM, which therefore have the same characteristic timescales.
First, there is the N\'eel switching. In this process, the magnetic moment of the MNP changes direction by overcoming the energy barrier $KV_c$, defined by the magnetocrystalline (or shape) anisotropy of the particle, with strength given by the anisotropy constant $K$ and the core volume $V_c$. 
In a suspension, the particles can also rotate as a whole. This process is called called Brownian rotation. In contrast to the N\'eel process, here the magnetization is fixed in the frame of the particle, but rotates (together with the particle) in the detectors' frame. The N\'eel fluctuation time $\tau_{\mathrm{N}}$ and the Brownian fluctuation time $\tau_{\mathrm{B}}$ are given by
\begin{equation}
  \tau_{\mathrm{N}} = \tau_0 \exp\left(\frac{KV_c}{k_{\mathrm{B}}T}\right)
  \quad
  \mathrm{and}
  \quad
    \tau_{\mathrm{B}} = \frac{3\eta V_{\mathrm{h}}}{k{_\mathrm{B}}T}
\end{equation}

where, $V_{\mathrm{c}}$ and $V_{\mathrm{h}}$ denote  the core and hydrodynamical volume of the particle, respectively. $\tau_0$ is the inverse of an attempt frequency, typically estimated on the order of $10^{-9}$ s, and and $\eta$ is the viscosity of the suspension. The effective fluctuation time is the inverse sum of the N\'eel and Brownian fluctuation times, 
$\tau_{\mathrm{eff}}=\frac{\tau_{\mathrm{N}} \tau_{\mathrm{B}}}{\tau_{\mathrm{N}}+\tau_{\mathrm{B}}}$
and is generally dominated by one contribution, depending on the particles' size.\\

Because of the inherent extremely small magnetic signals measured in TNM (down to a few fT), a deeper understanding of the TNM signal and its dependence on the sample configuration is required to improve this characterization technique. To this end, a theoretical framework is constructed to investigate different parameters that allow us to optimize our experimental setup. 
We describe and validate two ways of calculating the thermal noise power of a MNP ensemble and explain how they are used in the TNM measurements and simulations. Next, the dependence of the noise power on the number of particles and their volume is studied. Using the numerical framework, an optimized sample geometry for the experimental setup is proposed and subsequently fabricated. It is shown that it allows us to increase the measured noise power by a factor 3.5 while at the same time more than halving the necessary sample volume. The experimental and computational sensitivity profile, i.e. a noise power profile as a function of sample the distance from the detector, are compared. 

\section*{Methods}
\subsection*{Thermal Noise Magnetometry}
\label{Subsec:thermal_noise_magnetometry}
In Thermal Noise Magnetometry, fluctuations in the magnetic flux density of a nanoparticle ensemble are measured, which are caused by changes in the direction of the particles' magnetic moments due to thermal energy in the system. The magnetic flux density $B^{\lambda}$, resulting from the nanoparticle sample and evaluated at a certain point in space in a certain direction, is considered a stochastic variable. $B^{\lambda}$ is a superposition of all the magnetic fields attributed to the magnetic moments of the different particles in the sample, where $\lambda$ is one specific realization of the magnetic moments' phase space $\Omega$. 
Because the magnetic moment of every particle is free to move in any direction on the unit sphere, $\Omega$ is infinite and uncountable. $B^{\lambda}$ is described by the probability density function $P_B(x)$. Any statistical average of a function $f$ on the nanoparticle ensemble is then calculated as:
\begin{equation}
    \langle f(B^{\lambda})\rangle=\int f(B^{\lambda})P_{B}(x)dx
\end{equation}
At nonzero temperatures, the magnetization direction of the nanoparticles changes over time. $B^{\lambda}$ is time dependent and thus a stochastic process.  We assume the thermal fluctuations in the magnetic signal of a nanoparticle ($B^{\lambda}(t)$) ensemble to be stationary and ergodic: the statistical ensemble averages do not change over time and they can be set equal to the time averages:
\begin{equation}
    \langle f(B^{\lambda})\rangle=\overline{f(B^{\lambda}(t))}=\lim_{\substack{T\to\infty}}\left[\frac{1}{T}\int_{0}^{T} f(B^{\lambda}(t))dt\right]
\end{equation}
For stationary and ergodic processes, the bridge between theory (what can be calculated) and experiments (what can be measured) can readily be made. Whereas, the ensemble averages (the average of $B^{\lambda}(t^*)$ over the configurations $\lambda$ at fixed time $t^*$) are used for the theoretical calculations, the time averages, i.e. the average of one realization $\lambda^*$ of the stochastic process $B^{\lambda^*}(t)$ over time, are much more practical for experiments\cite{Refregier2004}.\\

There are two further important expressions for the stochastic process $B^{\lambda}(t)$. First, its autocovariance is given by
\begin{equation}
    \Gamma_{BB}(t_2-t_1)=\langle B^{\lambda}(t_1)B^{\lambda}(t_2) \rangle-m_x(t_1)m_x(t_2)=\Gamma_{BB}(\tau)=\langle B^{\lambda}(t_1)B^{\lambda}(t_1+\tau) \rangle-m_x^2,
\end{equation}
where $m_x(t)$ is the statistical average and the second equality holds for stationary processes. Secondly, its Power Spectrum is defined as 
\begin{equation}
    S_b(f)=\lim_{T\to\infty} \frac{1}{T}\int_{0}^{T}\langle |B^{\lambda}(t)\exp(-i2\pi ft)dt|\rangle=\frac{   (4\tau_{\mathrm{eff}})^{-1}   }{(\pi f)^2+(2\tau_{\mathrm{eff}})^{-2}}
    \label{eq: lorentzian}
\end{equation}
where the second equality holds for a monodisperse magnetic nanoparticle ensemble \cite{Machlup1954}.\\
\subsection*{Noise Power calculation}
\label{Subsec:noise_power_calculation}
The TNM signal can be quantified in the thermal noise power of the magnetic nanoparticles in the sample. We elaborate on two different methods to calculate this noise power. These methods are quantitatively compared for TNM in the Results and Discussion.
\subsubsection*{Method 1: configuration average}
 The instantaneous noise power of the stochastic process $B^{\lambda}(t)$ is given by its variance
\begin{equation}
    P_b(t)=\langle B^{\lambda}(t)B^{\lambda}(t)\rangle 
\end{equation}
which is independent of $t$ for stationary processes\cite{Papoulis2002,Refregier2004}. Calculating the statistical average is not possible without knowing the probability density function. The law of large numbers can therefore be used in order to approximate the statistical average:\\
\begin{equation}
    \langle (B^{\lambda})^2\rangle=\lim_{\substack{n\to\infty}}\frac{(B^{\lambda(1)})^2+(B^{\lambda(2)})^2+...+(B^{\lambda(n)})^2}{n}\label{large_numbers}
\end{equation}
If $n$ is large, the average of the squared magnetic flux density from $n$ random configurations $\lambda$ of $\Omega$ is a good approximation to estimate the magnetic noise power.\\

\subsubsection*{Method 2: PSD integration}
It is generally cumbersome or impossible to prepare a large number of uncorrelated experiments next to each other. Often, there is only one realization $\lambda$, which is evaluated over time. Measuring the noise power according to Eq. (\ref{large_numbers}) as a statistical average is thus experimentally unfeasible. By assuming that the process is stationary and ergodic, the ensemble averages can be replaced by time averages. Conventiently, the same assumptions also allows us to use the Wiener–Khinchin theorem\cite{Wiener1930,Khintchine1934}, which states that the autocovariance and the power spectral density are each other Fourier transforms.
\begin{equation}
    \Gamma_{BB}(\tau)=\int_{-\infty}^{\infty}S_b(f)\exp(2\pi i f\tau)df
\end{equation}
Because the the noise power equals the autocovariance for $\tau = 0$: $P_x=\Gamma_{XX}(0)$ , we can write:
\begin{equation}
    P_b=\Gamma_{BB}(0)=\int_{-\infty}^{\infty}S_b(f)df
\end{equation}
which means that the total noise power is given by the integral of the Power Spectral Density over the full frequency range.\\

Note that next to the noise power and the Power Spectral Density $\bigl($with units $fT^2$ and $\frac{fT^2}{Hz}$ respectively$\bigr)$, the noise amplitude and the Amplitude Spectral Density or the RMS Spectral Density $\bigl($with units $fT$ and $\frac{fT}{\sqrt{Hz}}$ respectively$\bigr)$ are also often used. Because the latter quantities are the square roots of the former, they both contain the same information. In this paper, we will work with the noise power and Power Spectral Density.\\

\subsection*{Thermal Noise Magnetometry experiments}
\label{Sec:TNM_EXP}
The Thermal Noise Magnetometry experiments were performed in the 6 channel MRX device at the Physikalisch-Technische Bundesanstalt in Berlin\cite{Ackermann2007}. The setup consists of 6 SQUID sensors with rectangular pickup coils, which are operated inside an integrated cylindrical superconducting magnetic shield. Through a warm bore with a diameter of 27 mm, the MNP sample can be positioned close to the SQUID sensors (typical distance 23.5 mm). For our measurements we only used one sensor. Iron oxide particles with an iron concentration of 1.214 mmol/L were used as an MNP sample. This nanoparticle system is similar to the one characterized in Ref. \cite{Leliaert2017} and was found to display a sufficiently high signal to allow measurements on a dilution series.\\
 
In each TNM measurement, we recorded 9,000 sampling frames of the time signal $B^{\lambda}(t)$ with a sample frequency $f_s$= 100 kHz. Each frame consist of $N=5,0000$ sample points. The power spectral density of each frame was then calculated in the following way:
\begin{enumerate}
  \item To overcome the problem of spectral leakage, each time signal $x_i'$ is first multiplied with the Hann window function $w_i$ ($i=0...N-1$): $x_i=x_i'\cdot w_i$. The fast Fourier transform (FFT) is then applied on $x_i$ to yield FFT($x_i$)=$a_i$. Because only positive frequencies are to be considered, the Fourier amplitudes belonging to the negative frequency part are mapped to the positive frequency part. The Fourier amplitudes $a_j$ belonging to this positive frequencies gain a factor of two\cite{Cerna2000}.
  \item The Power Spectral Density is then computed as
\begin{equation}
    S_j=\frac{2\cdot |a_j|^2\cdot \eta^2}{f_s\cdot N}
\end{equation}
$\eta$ is a factor accounting for the window usage, and is equal to $\frac{2}{\sqrt{1.5}}$ for the Hann window. For a detailed explanation of the discrete PSD computation using windows we refer to Ref. \cite{Heinzel2002}.\\
\end{enumerate}
The different PSDs are then averaged over the 9,000 frames. The uncertainty on the PSD values was calculated from the standard deviation of the 9,000 frames.\\
\subsubsection*{Noise power and uncertainty calculation}
The noise power was calculated by integrating the resulting PSD over the frequency domain following method 2. Two contributions for the uncertainty $\Delta P$ on the power were taken into account. Firstly, we account for the uncertainty coming from the calculated uncertainties on the PSD \cite{Woolliams2013}. Secondly, small amplitude variations in the background noise power were taken into account by measuring an empty sample and analysing its data the same way as if it would have been a MNP sample.\\

\subsection*{Thermal Noise Magnetometry simulations}
\label{Sec:TNM_SIM}
The thermal noise magnetometry simulations were performed with Vinamax \cite{Leliaert2015a}: a macrospin simulation tool for magnetic nanoparticles, which was recently extended with the ability to simulate Brownian rotations of the particles\cite{coene2020}. The shape of the sample holder was defined to closely resemble those used in the experiments and the magnetic field was evaluated in a rectangular detector with the same dimensions as the SQUID pickup coil.\\
\begin{figure}[h!]
  \centering
    \includegraphics[width=1\textwidth]{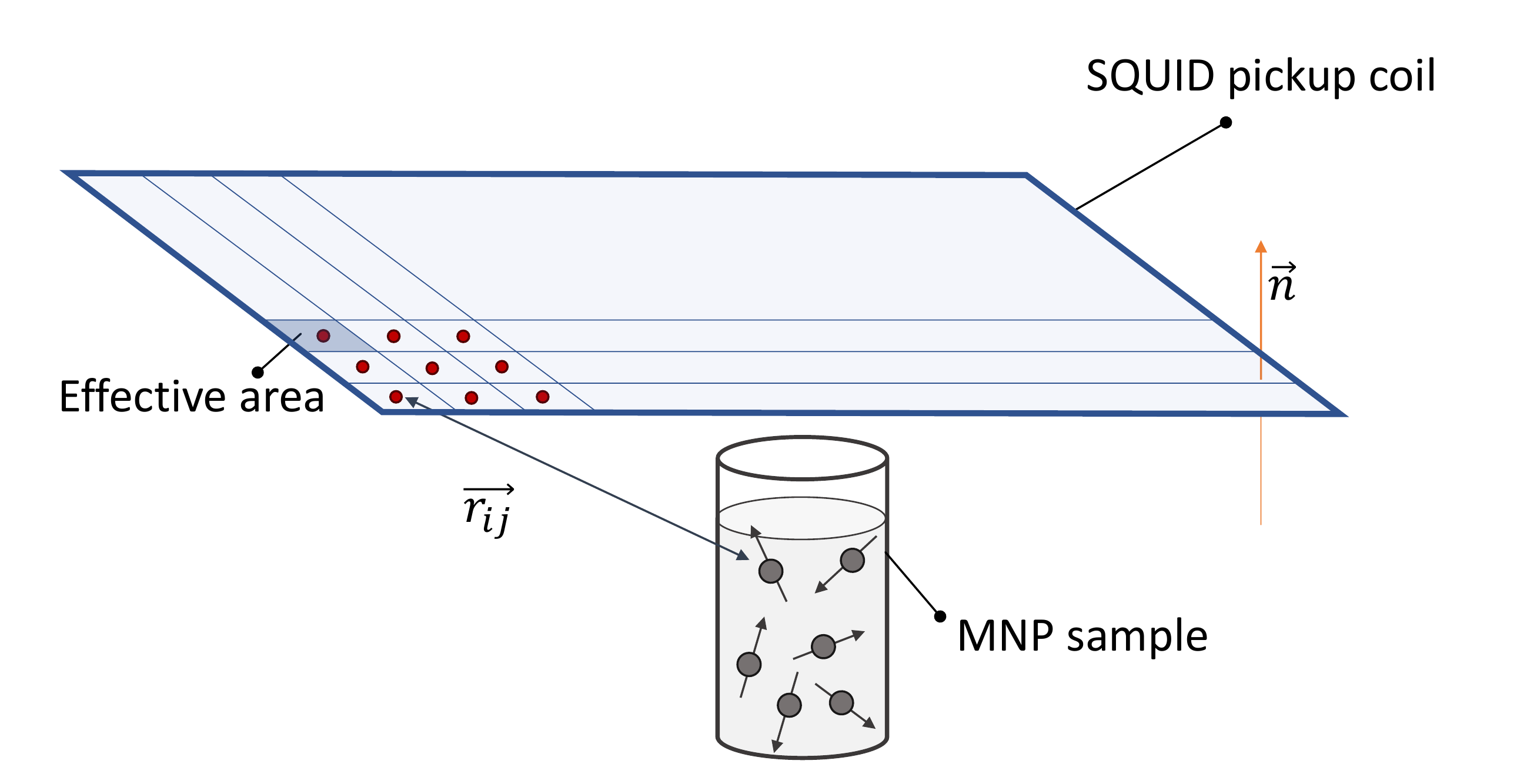}
    \caption{Schematic representation of the macrospin simulations modeling a thermal noise magnetometry experiment. The particles $i$ and the detector points $j$ are displayed, connected through the vector $\vec{r_{ij}}$. The detector only measures the magnetic field in the direction $\vec{n}$, the normal on the detector plane. Each detector point contributes for a certain effective area, which is sufficiently small to ensure a homogeneous field in each area.}
    \label{Fig:vectors}
\end{figure}
\subsubsection*{Magnetic flux density in detector}
For the calculation of the magnetic field density $B$, the detector was discretized in $N$ points (with index $j$) and the magnetic flux density $\vec{B_{ij}}$ of every simulated particle $i$ was evaluated in point $j$ as a dipole-dipole field (see Fig. \ref{Fig:vectors}):
\begin{equation}
    \vec{B_{ij}} = \frac{\mu_{0}}{4\pi}\left(\frac{3 \vec{r_{ij}}(\vec{m_{i}}\cdot\vec{r_{ij}})}{|\vec{r_{ij}}|^5}-\frac{\vec{m_{i}}}{|\vec{r_{ij}}|^3}\right)
\end{equation}

Here, $\vec{r_{ij}}$ denotes the vector between particle $i$ and detector point $j$ and $\vec{m_{i}}$ is the magnetization of particle $i$. The detector is only sensitive to fields in the direction perpendicular to its surface:
\begin{equation}
B_{ij} = \vec{n}\cdot\vec{B_{ij}}
\end{equation}
where $\vec{n}$ is the normal on the detector plane. The magnetic flux density in each point $j$ of the detector is the sum of the contributions from each particle:
\begin{equation}
    B_j = B_{i_1j} + B_{i_2j} + ...
\end{equation}
 and accounts for the magnetic flux (in Wb) through the \textit{effective area} of this point: the area of the detector which it represents. For equidistant detector discretization, the total flux density in the full detector is then calculated by taking the average of the magnetic flux densities in the different detector points $j$.\\

\section*{Results and discussion}
\subsection*{Validation of Noise power calculation methods}
\begin{figure}[h!]
  \centering
  \includegraphics[width=1\textwidth]{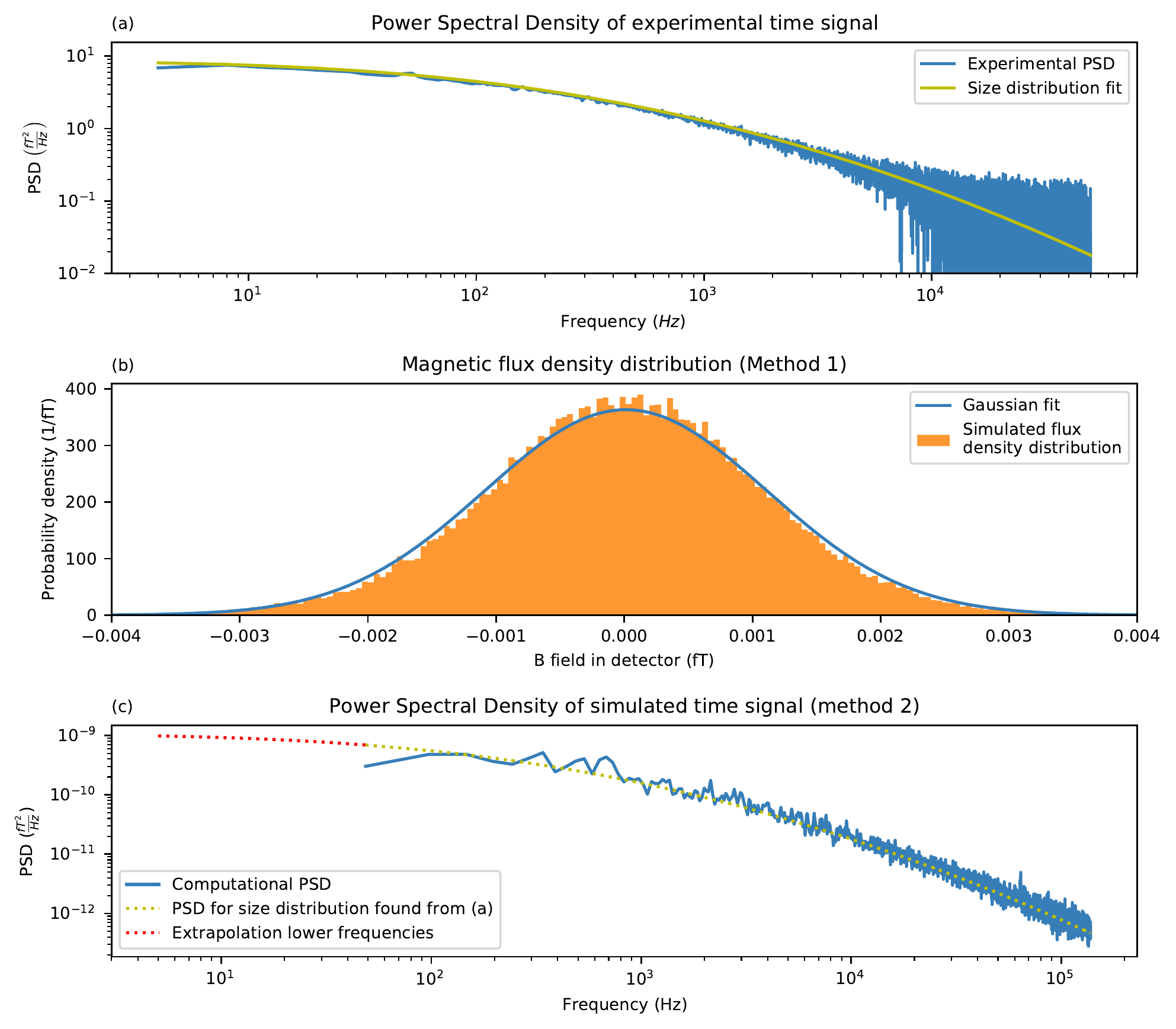}
    \caption{Power Spectral Density of the experimental time signal and fitted lognormal distribution (a). This size distribution is subsequently used in the simulations to validate the two noise power calculation methods. Panel (b) displays the thermal magnetic field distribution in the detector of 100,000 simulated configurations (method 1). The distribution of the stochastic magnetic signal deviates slightly from a Gaussian distribution. The Power Spectral Density of the computational time signal generated with Vinamax is displayed in (c), which is used to calculate the noise power following method 2. An extrapolation of calculated PSD in the lower frequency range is plotted, using the PSD form of the assembling size distribution.}
    \label{Fig:Power_calculation}
\end{figure}

We now compare the noise power obtained using both methods presented above. To this end, we fit a volume-squared weighted lognormal particle size distribution (see appendix) to the experimentally measured spectrum of the iron oxide MNPs in a cylindrical sample holder, shown in Fig. \ref{Fig:Power_calculation} (a). 
This distribution, yielding parameters $\mu=46.14$ nm and $\sigma=0.63$, is subsequently used to simulate a MNP ensemble using Vinamax. Each simulations contained 10,000 particles. Only Brownian switching was taken into account, and we used the relevant material parameters of iron oxide (saturation magnetization $M_s=$400 kA/m) and the suspension parameters for water (viscosity $\eta$ = 1 mPa$\cdot$s). 
Particles with a diameter larger than 10 nm have been left out of the simulation, since their characteristic spectral density is constant in the considered frequency range.\\

The simulations naturally lend themselves to generate a large amount of randomly chosen configurations $\lambda$, which allows one to calculate the noise power using \emph{method 1}. Generating 80,000 sample configurations, the noise power was calculated by Eq. (\ref{large_numbers}) to be $(1.21 \pm 0.03)\cdot10^{-6}$ fT$^2$. The distribution of the magnetic flux density $B^{\lambda}$ in the detector is shown Fig. \ref{Fig:Power_calculation} (b). As expected for the lognormal size distribution, the magnetic noise appears to be slightly non-Gaussian.\\

It is also possible to generate a time series of the MNP dynamics in Vinamax, similar as the time signal recorded in experiments. A video, displaying the time dependent magnetic field in the detector plane is shown in the Supplementary Material. The simulated time signal allows one to estimate the noise power following method 2, which can quantitatively be compared to the noise power found by method 1. The averaged power spectral density from such a simulated signal is shown in Fig. \ref{Fig:Power_calculation} (c). Note that its amplitude does not match the amplitude of the experimental PSD in panel (a), since only 10,000 particles were considered in the simulation. The analytic expression of the PSD for the corresponding size distribution is plotted as well and an extrapolation towards the lower frequency part is made. The integration of the combined PSD (i.e. the simulated and extrapolated part up to 0.01 Hz) yields a noise power of $(1.19 \pm 0.03)\cdot10^{-6}$ fT$^2$.\\

The noise power resulting from the two methods coincide within their uncertainty intervals. Both methods yield equal noise powers, and can thus be interchanged for TNM. It is clear that method 1 is more convenient for simulations. Independent sample configurations are easily generated, while time signals consume more calculation time to cover the low frequency range of the PSD. For the experiments however, the noise power calculations naturally follow method 2.

\subsection*{Noise power dependence on particle volume and number of particles}
Now that the TNM noise power can be calculated efficiently both for simulations and experiments, the simulations can be used to improve the experiment. The number and volume of particles are two straightforward parameters which can be adjusted experimentally and whose influence on the signal strength are well understood in other characterization methods. The stochastic nature of the TNM experiment makes their impact however less intuitive. Here, we investigate the dependence of the thermal noise power on the number of particles and their volume, both for a variable and a fixed total iron concentration. 


\subsubsection*{Variable iron concentration}
\begin{figure}[h!]
  \centering
    \includegraphics[width=1\textwidth]{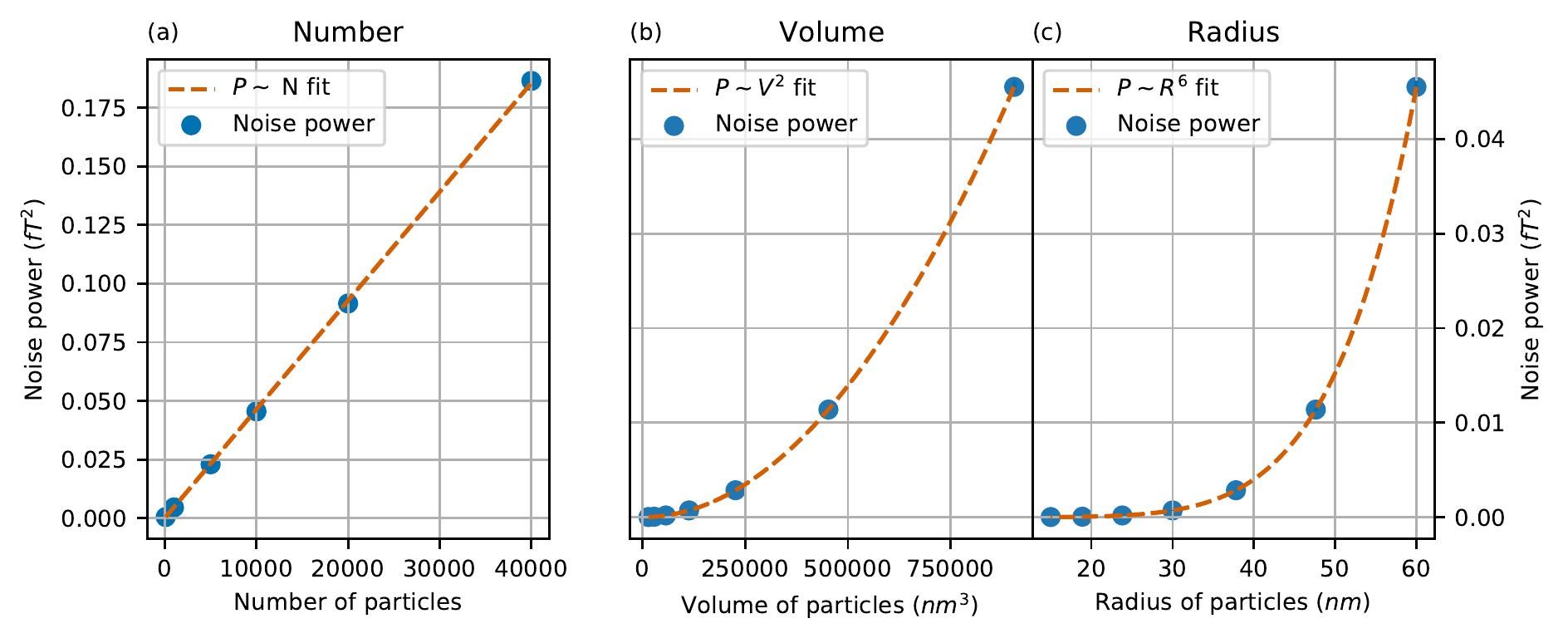}
    \caption{Noise power for a variable total iron amount as a function of (a) the number of particles (with a fixed volume) and (b) the volume of the particles (with a fixed number of particles). A cylindrical sample holder with monodisperse MNPs has been used. The noise power scales linearly with the number of particles and quadratically with their volume. The latter is also shown as function of particle radius in (c), based on the same dataset. Note that uncertainties have been calculated, but are too small to be seen.}
    \label{Fig:Variable_Fe}
\end{figure}

Fig. \ref{Fig:Variable_Fe} shows the dependence of the noise power on the number of particles obtained by simulations (with a fixed volume, equivalent to a radius of 60 nm) in panel (a) and the  effect of the particle volume (for a fixed number of 10,000 particles) in panel (b). Note that in both cases, the total iron amount is not constant. A cylindrical sample holder with monodisperse MNPs has been used for the simulation.

The noise power scales linearly with the number of particles, and quadratically with their volume. These power laws can be explained as follows. 
The magnetic nanoparticles in the sample are noise sources, which means that adding an additional particle to the ensemble can either result in a positive or a negative contribution to the magnetic field in the detector, depending on the orientation of the particle's magnetic moment. On average, when doubling the number of particles the flux density in the detector will only increase by a factor $\sqrt{2}$, and the noise power will scale linearly with the number of particles. This contrasts deterministic processes, e.g. MRX measurements, in which all particles are aligned by an externally applied field\cite{Wiekhorst2012}, where the signal amplitude scales linearly, and hence the power scales quadratically with the amount of particles.\\

Increasing the particles' volume while the number of particles stays constant, increases the TNM signal as if it would have been a deterministic signal. No extra noise sources are created, and the larger magnetic moments results in a quadratic increase of the noise power, as shown in Fig. \ref{Fig:Variable_Fe} (b) and (c).\\

\begin{figure}[h!]
  \centering
    \includegraphics[width=1\textwidth]{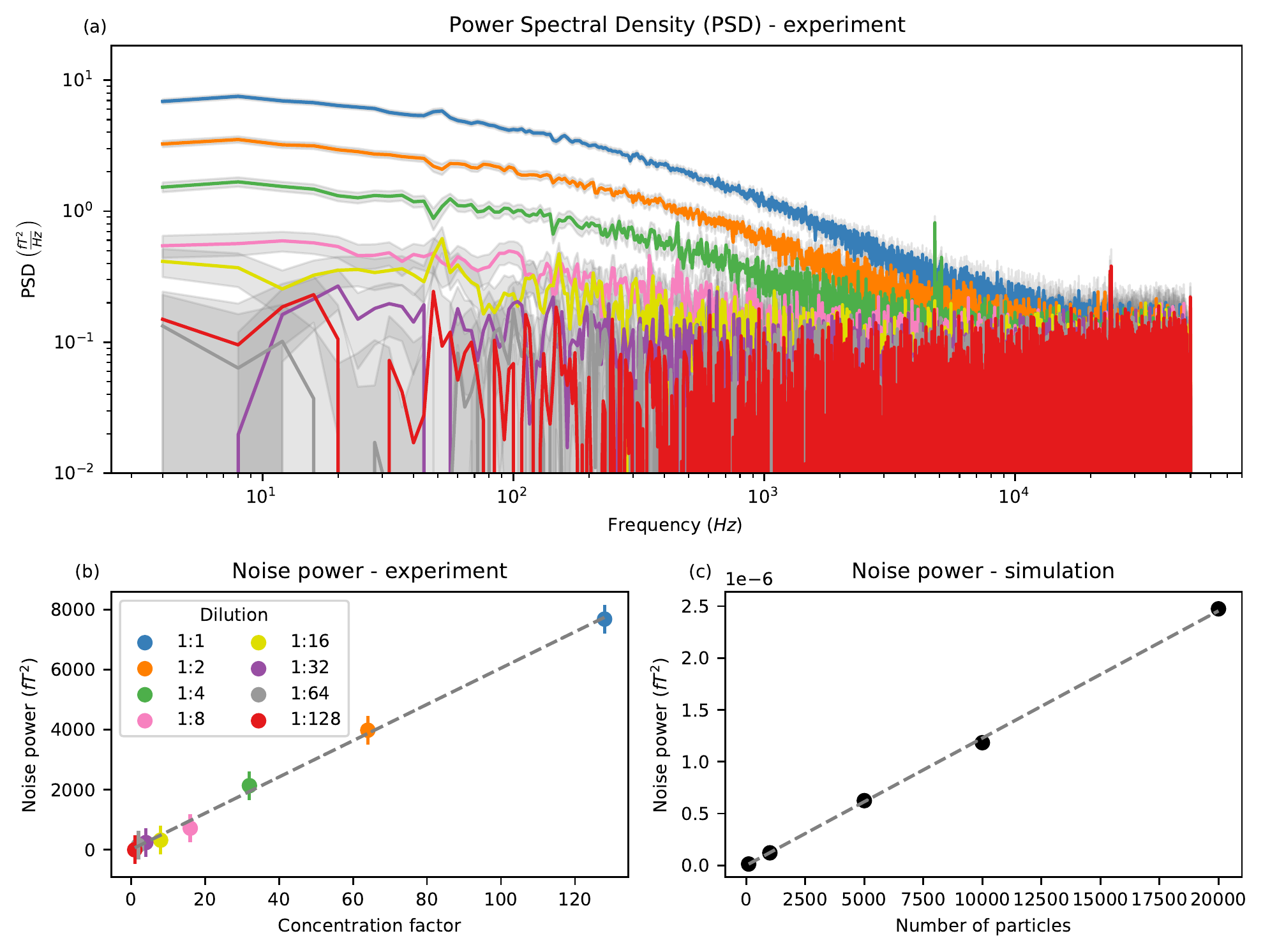}
    \caption{Experimental confirmation of the linear dependence of the noise power on the number of particles. (a) The measured Power Spectral Density of the dilution series and  (b) their integrated noise power as a function of the concentration factor, i.e. the inverse dilution factor. Panel (a) and (b) share the same legend. (c) shows the simulated noise power as a function of number of sample particles, where the particles' size distribution is chosen to resemble the experiment.}
    \label{Fig:Dilution_series}
\end{figure}

The linear dependence of the noise power on the number of particles was experimentally confirmed, as shown in Fig. \ref{Fig:Dilution_series}. We measured a dilutions series of the iron oxide MNPs in a water suspension. The power spectral density (PSD) of each dilution and their integrated noise power are shown in panels (a) and (b), respectively. The peaks observed at 50 Hz and a few higher frequencies are artefacts and have not been taken into account when calculating the total noise power. To make a consistent quantitative comparison with the computational results, the simulation of Fig. \ref{Fig:Variable_Fe} (a) was repeated with the lognormal size distribution parameters found from the fit in Fig. \ref{Fig:Power_calculation} (a). The results are shown in Fig \ref{Fig:Dilution_series} (c). The total amount of particles in the sample can then be estimated from a quantitative comparison between the computational results of Fig. (b) and (c). Taken into account the linear dependence on the amount of particles, the non-diluted sample (1:1) is estimated to  contain about $6\cdot10^{13}$ particles.\\

\subsubsection*{Fixed iron concentration}
\begin{figure}[h!]
  \centering
    \includegraphics[width=1\textwidth]{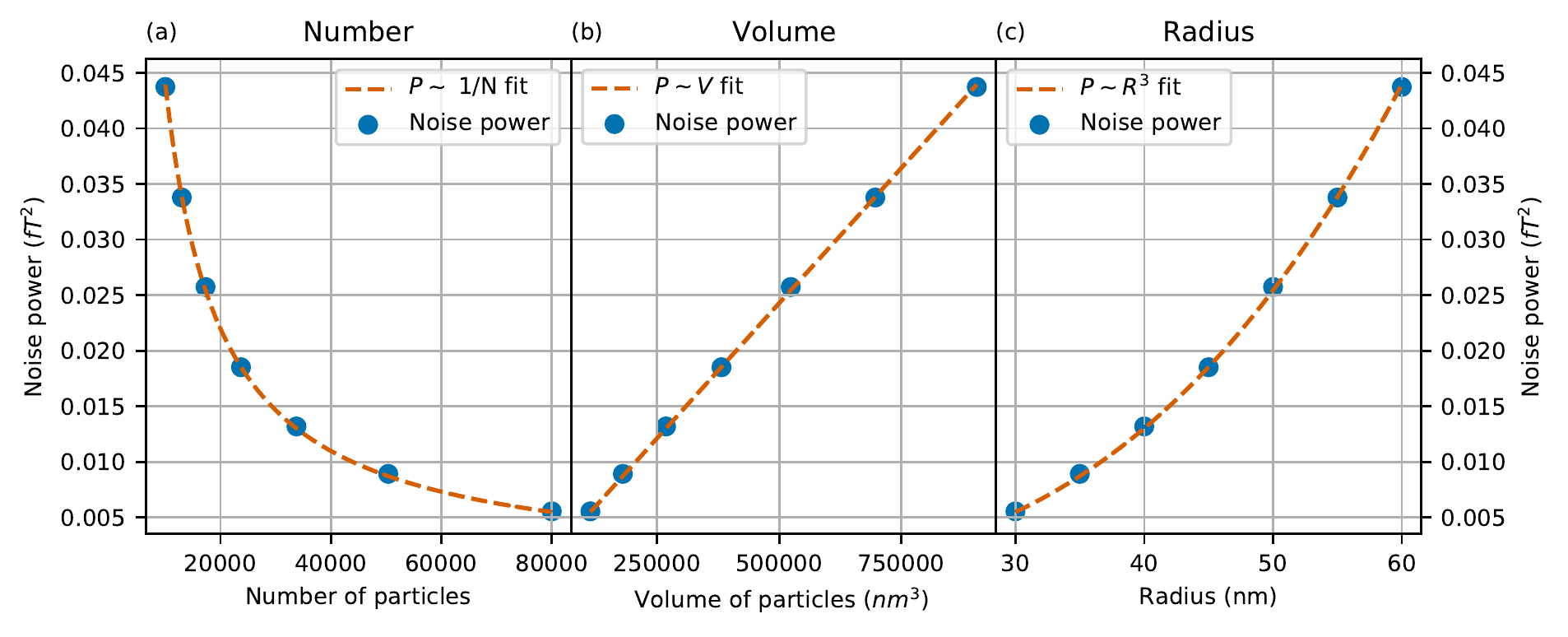}
    \caption{Noise power for a fixed total iron amount (fixed total volume) as a function of (a) the radius of the particles, (b) the number of particles and (c) the volume of the particles. A cylindrical sample holder with monodisperse MNPs has been used. All subfigures show the same dataset, but on different axes.}
    \label{Fig:Fixed_Fe}
\end{figure}

Let's turn our attention to the case in which the total iron amount in the sample remains constant. Fig. \ref{Fig:Fixed_Fe} shows the noise power of a
simulated particle ensemble where the number of particles $N$ and their volume $V_p$ are varied according to $V_{tot}=N\cdot V_{p}$ for a fixed $V_{tot}$. All panels display the same dataset and show that the noise power depends linearly on the particle volume and inversely on the number of particles.  This is readily understood from the power laws described in the previous paragraph: for a variable particle amount, when increasing the amount of particles by a factor $X$, the noise power also increases by the same factor. However, because the total iron concentration is fixed, this means that the particle volume is reduced by $X$, which on its own would lead to a reduction of a factor $X^2$ of the noise power. Together, they result in an inverse linear, and a linear relation between the noise power and the particle number and volume, respectively, for a fixed total iron concentration. This means that in a dynamic measurement where we track the signal over time, clustering of the particles could be observed as an increase in noise power due to a decrease in the number of particles in favour of larger particle volumes. 

\subsection*{Design and validation of an optimized sample holder geometry}
Next to investigating the noise power dependence on parameters of the samples content, one can also improve the samples geometry in favor of the TNM signal. The magnetic field of a (point) dipole decays as $1/r^3$ with  distance $r$. Therefore, the distance of the MNPs from the detector has a major influence on the amplitude of the TNM signal and the sample is ideally positioned as closely to the detector as possible. However, the geometry of the setup and the nonzero volume of the sample impose certain constraints, resulting in a trade off between signal strength and amount of material. Often, symmetrically shaped sample holders (cylinders, cones) are employed for measurements. These are easy to manufacture, but signal might be lost due to the chosen geometry. Therefore, we calculate a geometry and design a sample holder that enhances the TNM signal in the given setup.\\

First, the available sample space in the warm bore was discretized into regularly spaced grid of voxels, spaced 500 $\mu$m apart. For each voxel, the hypothetical contribution of one MNP, located inside voxel, to the total noise power measured in the detector was calculated. By sorting the sampled voxels by their impact on the measurement result, a ranking of the most effective points was made. Selecting those voxels with the highest impact then defines an optimized sample geometry for this specific volume.\\

\begin{figure}[h!]
  \centering
    \includegraphics[width=1\textwidth]{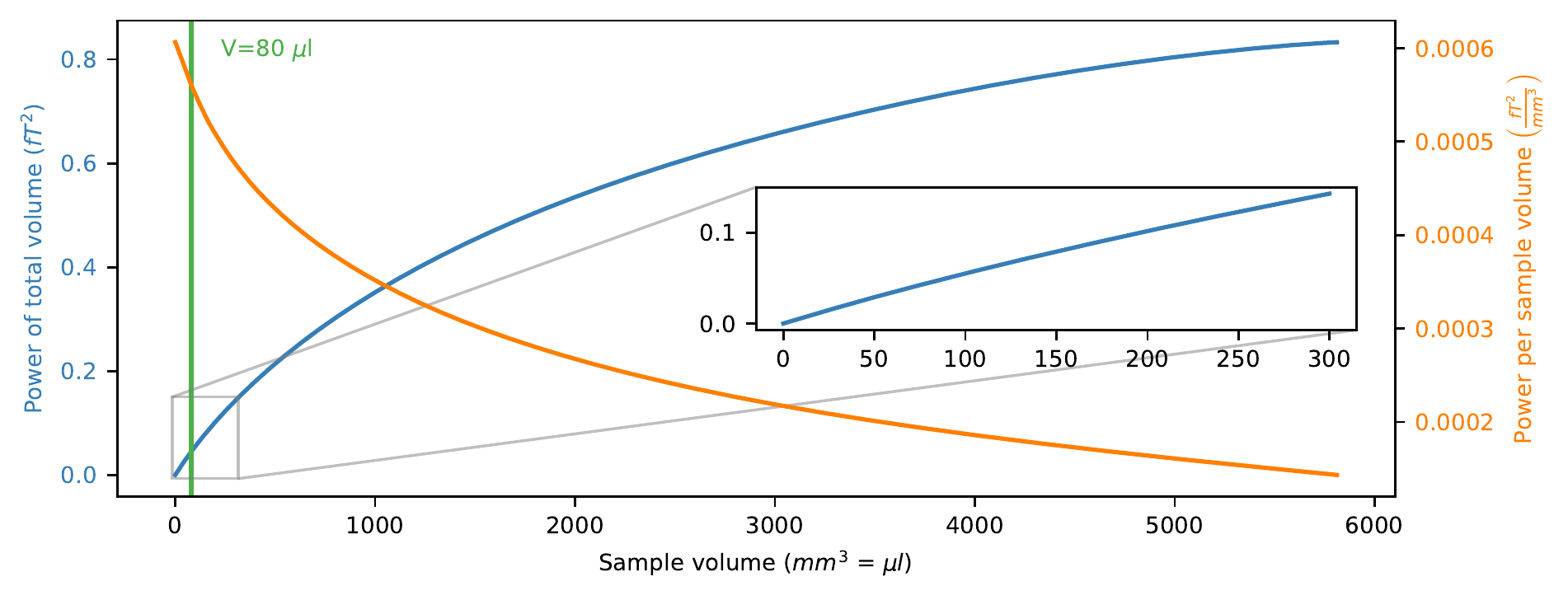}
    \caption{Calculated noise signal (blue) and noise power gain per added volume element (orange) as a function of the sample volume. A quasi linear regime for the noise power is visible for common sample volumes (up to 300 $\mu $l). The geometry of the sample of 80 $\mu $l was used for the construction of the experimental optimized sample holder.}
    \label{Fig:Sample_volume}
\end{figure}

Fig. \ref{Fig:Sample_volume} shows the total noise power as a function of the sample volume for the optimized sample geometry (blue curve). The total noise power as function of sample volume is a monotonically increasing function. However, the rate at which it increases decreases for larger volumes because points far away from the detector contribute negligibly to the TNM signal.

This is also reflected in the power per sample volume, as plotted in orange in Fig. \ref{Fig:Sample_volume}. The most ``efficient'' sample, corresponding to the highest noise power per sample volume, would consist of only a single MNP. Although the fluctuations of individual particles can be captured\cite{piotrowski2014magnetic}, the noise power would be insufficient to detect in our setup, which is aimed at characterizing magnetic nanoparticle samples containing over 10$^9$ particles.\\

\begin{figure}[h!]
  \centering
    \includegraphics[width=0.6\textwidth]{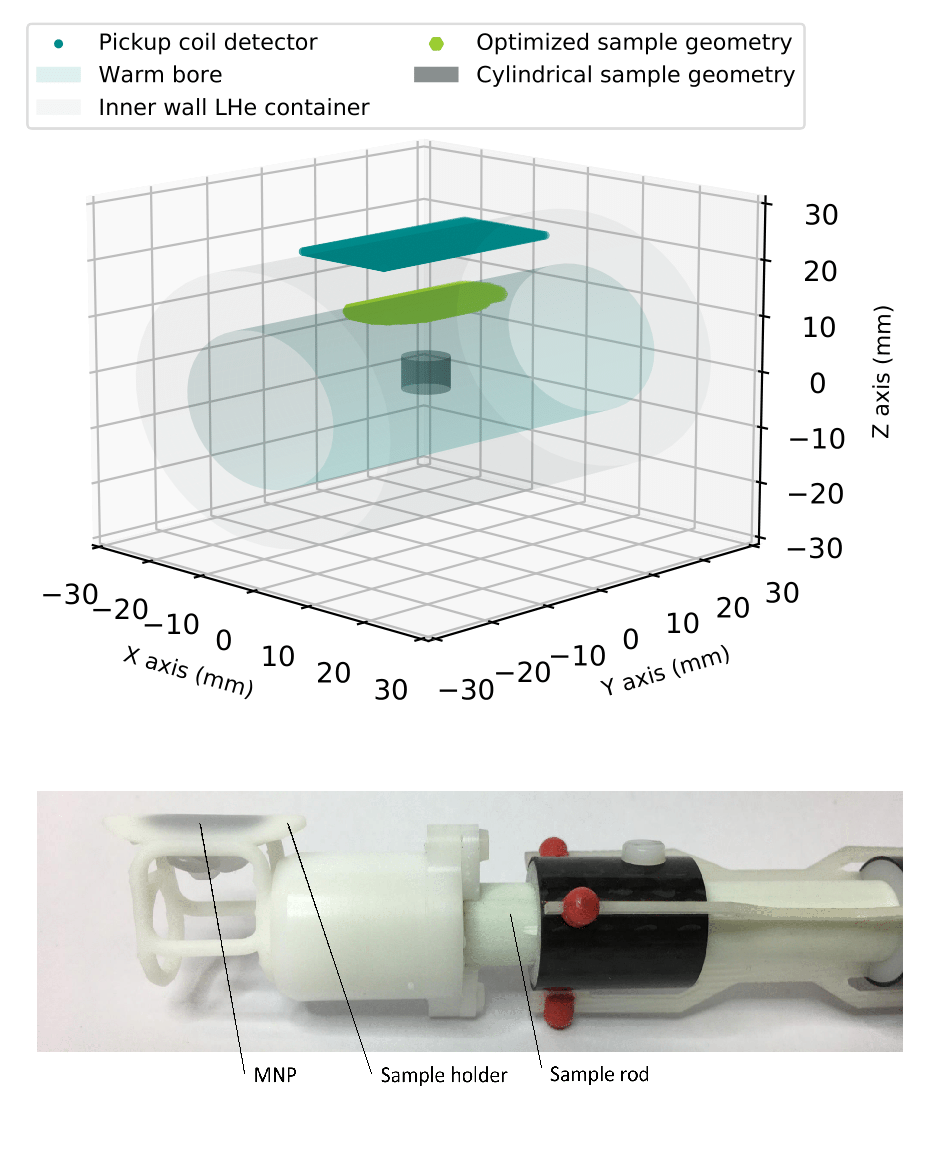}
    \caption{Upper: Calculated optimized sample geometry for a volume of 80 $\mu $l in the experimental setup. The cylindrical sample is also plotted for comparison. Lower: Picture of the printed sample holder, used in the TNM experiments. The sample holder is filled with magnetic nanoparticels. The sample rod serves to place the sample at the right position in the warm bore.}
    \label{Fig:Sample_geometry}
\end{figure}

Because the smooth decrease of the orange noise power per volume curve, there is no clear cutoff value for the sample volume. Instead, bearing in mind the cost of the material and practicality of the measurements, we chose to optimize our experimental sample holder geometry for a volume of 80 $\mu $l. The resulting shape is shown in Fig. \ref{Fig:Sample_geometry}. These coordinates were then used to construct a sample holder for experimental use. The 3D additive manufactured experimental sample holder is displayed on the lower part of Fig. \ref{Fig:Sample_geometry}.

One way to further increase the signal per sample volume would be to use all 6 detectors of the setup. Instead of placing the entire sample as close as possible to one detector, the volume could be divided into 6 subvolumes divided over the 6 detectors. However, because of the nearly linear $P(V)$ curve for small volume changes $V$ (see the inset of Fig. \ref{Fig:Sample_volume}), such a construction would barely increase the noise power for these volumes and would significantly increase the sample preparation time. Therefore, only one detector was used for the construction of our optimized sample holder.\\

\subsubsection*{Experimental validation}
To validate the 80 $\mu$l optimized sample holder, the dependence of the noise power on the samples' distance from the detector is investigated and compared to the original cylindrical sample holder. Because the sample can only move in one direction (the y-direction in Fig. \ref{Fig:Sample_geometry}), we construct a power profile along this axis in the setup. This profile is further referred to as the sensitivity profile.\\

\begin{figure}[h!]
  \centering
    \includegraphics[width=1\textwidth]{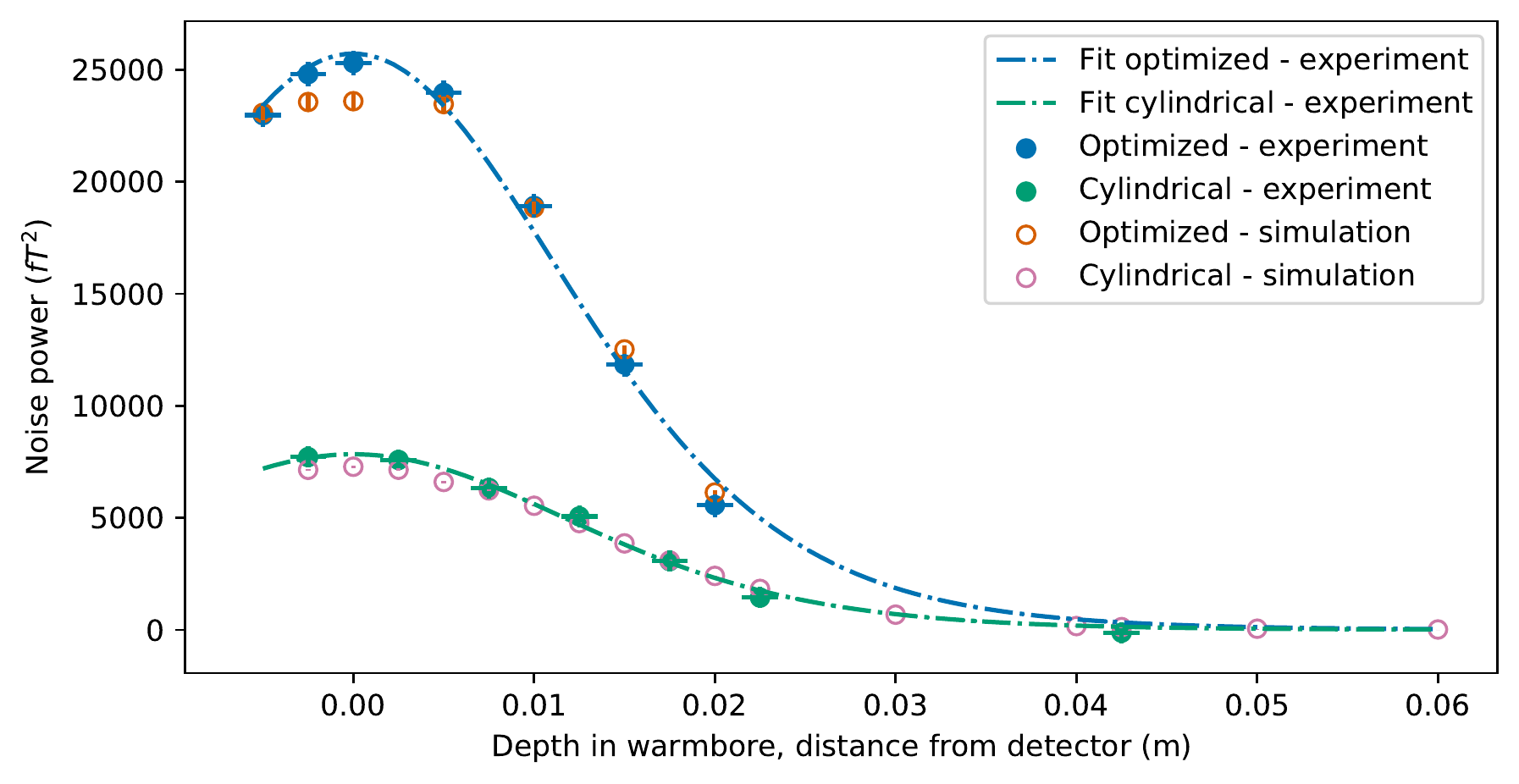}
    \caption{Sensitivity profile of the cylindrical and the optimized sample holders, both measured experimentally and computationally. By varying the y-coordinate of the sample holders, thereby altering the distance to the detector, a noise power sensitivity profile is constructed as a function of the position of the sample holder in the warm bore. The values of the simulations are scaled in order to match the experimental values. The experimental curves are fitted with the dipole-dipole sensitivity profile (Eq. (\ref{Eq:dipole})).}
    \label{Fig:Depth_profile}
\end{figure}

The measured noise power as function of position, as well as the simulated counterpart (based on 10,000 sample configurations with the lognormal size distribution parameters from Fig \ref{Fig:Power_calculation} (a) and rescaled to match the experimental noise power) are presented in Fig. \ref{Fig:Depth_profile}. The lognormal size distribution from the fit in Fig. \ref{Fig:Power_calculation} (a) has again been used to model the experiment as well as possible.\\

The measured and simulated sensitivity profiles are similar, although the experimental data typically display slightly sharper peaks, i.e. higher values around 0 and lower values in the tails at larger distances. The ratio between the experimental and the computational noise power (at distance zero) is 1.07 and 1.06 for for the optimized and the cylindrical sample holders, respectively.
This small difference is attributed to a systematic inaccuracy, e.g. a small difference between the theoretical and the actual length of the detector in the experiment. Nonetheless, the good agreement between our experimental en simulated results validates the computational framework used to simulated the thermal noise of the magnetic particles.\\

As a guide to the eye, the experimental curves are fitted with a dipole-dipole sensitivity profile model:
\begin{equation}
    P_b=a \left( \frac{1}{(d^2+b^2)^3}+\frac{1}{(d^2+b^2)^5}\right) 
    \label{Eq:dipole}
\end{equation}
where $P_b$ is the noise power, $d$ the position in the warm bore, $b$ the distance between the sample and the detector and $a$ a weight factor (which could e.g. account for the amount of particles in the sample).\\
For the cylindrical sample holder, the dipole fit matches the data quite well. Indeed, the larger distance between the sensor and the narrow cylinder allows to approximate the sample as a point source. From the geometry shown in Fig. \ref{Fig:Sample_geometry} it is obvious that a simple point-point model between the detector and the optimized sample holder is not a good approximation.\\

The main conclusion that can be drawn from Fig. \ref{Fig:Depth_profile} is that our sample holder significantly improves the measurements: despite only containing 80 $\mu$l sample material (as compared to the 200 $\mu$l in the original cylindrical sample holder), the optimized sample holder yields an increase of the TNM signal by a factor of 3.5 as compared to the original cylindrical sample holder.

\section*{Summary and Conclusion}
We present a theoretical framework to examine the noise power properties of magnetic nanoparticles in Thermal Noise Magnetometry (TNM). 
The dependence of the TNM signal on basic, yet fundamental, parameters such as the number of particles and their volume has been studied, and the resulting power laws have been explained. We describe how we designed a sample holder geometry to maximize the noise power measured in our setup, and compare its noise signal to a cylindrical sample holder typically used for MRX measurements. We obtained an increase in noise signal by a factor of 3.5, while the volume of the sample decreased by more than half. Both the optimized sample holder and the computed sensitivity profiles validate our TNM simulations, showing that we can computationally model the TNM experiment. The theoretical framework allows us to predict and visualize aspects of TNM that then can be analyzed and studied experimentally. Our results therefore contribute to the further establishment of TNM as a reliable magnetic nanoparticle characterization method.

\bibliographystyle{unsrt} 

\section*{Acknowledgements}
This work was financially supported by the German research foundation (DFG), project "MagNoise: Establishing thermal noise magnetometry for magnetic nanoparticle characterization" FKZ WI4230/3-1.\\
This work was supported by the Fonds Wetenschappelijk Onderzoek (FWO-Vlaanderen) with a postdoctoral fellowship (JL).

\section*{Author contributions statement}
J.L. and F.W. defined the project. K.E., J.L., F.W. and M.L. conceptualized the study. K.E conducted the measurements and analysed the results. K.E. and J.L. performed the simulations. D.G. and K.E. designed and manufactured the optimized experimental sample holder. K.E., F.W., M.L. and J.L. discussed the results and prepared the manuscript. All authors reviewed the manuscript. J.L., F.W. and B.V.W. supervised the project.

\section*{Competing interests}
The authors declare no competing interests

\section*{Supplementary Material}
Video of the simulated thermal magnetic field in the detector. The nanoparticle ensemble consist of 10,000 monodisperse spherical nanoparticles with diameter 60 nm in the cylindrical sample holder. The pickupcoil in the 6 channel setup has an area 4 times smaller then the one displayed, which would be centered in the middle. (See file "Time variation detector.mp4")

\section*{Additional information}
To include, in this order: \textbf{Accession codes} (where applicable); \textbf{Competing interests} (mandatory statement). 

The corresponding author is responsible for submitting a \href{http://www.nature.com/srep/policies/index.html#competing}{competing interests statement} on behalf of all authors of the paper. This statement must be included in the submitted article file.

\appendix

\section*{Appendix: The lognormal size distribution}
In this paper, as is often the case for MNP\cite{kiss1999new}, we assume a lognormal size distribution of the magnetic nanoparticles.\\

We use the following definition for  the lognormal distribution, with variable $x$:
\begin{equation}
    P(x,\mu,\sigma^2)=\frac{1}{\sqrt{2\pi}\sigma x}\exp\left(\frac{-[\ln{(x/\mu)]^2}}{2\sigma^2}\right)
    \label{eq:lognormal}
\end{equation}
This distribution is fully determined by $\mu$ and $\sigma$, which correspond to the median and geometric standard deviation of the distribution, respectively.\\

The TNM Power Spectral Density consists of a (volume-squared) weighted sum of Lorentzian (see Eq. (\ref{eq: lorentzian})) curves, each corresponding to the noise spectrum of a single nanoparticle. This signal can be decomposed in a lognormal distribution of such curves. However, often we are not interested in this \emph{relaxation time} distribution, but in the corresponding \emph{diameter} distribution, which itself can be either number, volume or volume-squared weighted.

In this appendix, we derive and recall some useful properties of the lognormal size distribution in order to perform these transformations.

Note that there is a fundamental difference between transforming the volume and number weighted diameter distributions into each other, and the transformation of the (e.g. volume weighted) diameter distribution into the (e.g. also volume weighted) switching time distribution.

In the first transformation, we seek to describe two \emph{different} distributions as function of \emph{the same} quantity, i.e. diameter.
In contrast, the second transformations describes \emph{the same} distribution as function of \emph{different} quantities, i.e. diameter or switching time.

\subsection*{Transforming a number weighted distribution into a diameter weighted distribution and vice versa}
In Ref. \cite{ELHILO20122593}, it is shown for the lognormal size distribution how to transform a volume weighted distribution into a number weighted distribution. Here, we extend this result to a volume-squared weighted distribution:
\begin{equation}
D^6 \overbrace{P(D,\mu,\sigma^2)}^{\mathrm{number\,\,weighted}}dD\propto D^3\overbrace{P(D, \exp[\ln(\mu)+3\sigma^2],\sigma^2)}^{\mathrm{volume\,\,weighted}}dD\propto \overbrace{P(D, \exp[\ln(\mu)+6\sigma^2],\sigma^2)}^{\mathrm{volume-squared\,\,weighted}}dD\\
\end{equation}


\subsection*{Transforming a switching time distribution into a diameter distribution and vice versa}
\begin{equation}
P(\tau_B,\mu_\tau,\sigma_\tau^2)d\tau_B\propto P\left(D,\left[\frac{2k_\mathrm{B}T}{\eta\pi}\mu_\tau\right]^{1/3}
,\frac{\sigma_\tau^2}{9}\right)dD
\end{equation}
and
\begin{equation}
P(D,\mu_D,\sigma_D^2)dD\propto P\left(\tau_B,\frac{\pi\eta}{2k_\mathrm{B}T}\mu_D^3,9\sigma_D^2\right)d\tau_B
\end{equation}

Where we made use of the following properties of the lognormal distribution\cite{ELHILO20122593}:\\

Multiplication with a scalar $a$: if variable $x$ is lognormal distributed with mean $\mu$ and standard deviation $\sigma$, then $y=a\cdot x$ is lognormal distributed with mean $a\mu$ and standard derivation $\sigma$.
\begin{equation}
P(x,\mu,\sigma^2)dx=P(y,a\mu,\sigma^2)dy
\end{equation}
    
Exponentiation with a scalar $a$: if variable $x$ is lognormal distributed with mean $\mu$ and standard deviation $\sigma$, then $y=x^a$ is lognormal distributed with mean $\mu^a$ and standard derivation $a\sigma$.
\begin{equation}
    P\left(x,\mu,\sigma^2\right)dx= P\left(y,\mu^a, a^2\sigma^2\right) dy
\end{equation}

\end{document}